\documentclass[a4paper,twoside]{article}
\makeatletter
\newcommand*{\rom}[1]{\expandafter\@slowromancap\romannumeral #1@}
\makeatother
\usepackage{float}
\usepackage{epsfig}
\usepackage{subcaption}
\usepackage{calc}
\usepackage{amssymb}
\usepackage{amstext}
\usepackage{amsmath}
\usepackage{amsthm}
\usepackage{multicol}
\usepackage{pslatex}
\usepackage{apalike}
\usepackage{SCITEPRESS} 

\begin{document}

	\title{Statistical Inference of the Inter-sample Dice Distribution for Discriminative CNN Brain Lesion Segmentation Models}

	\author{\authorname{Kevin Raina \sup{1}\orcidAuthor{0000-0002-6240-9675}}
		\affiliation{\sup{1}Department of Mathematics and Statistics, University of Ottawa, Ontario, Canada}
		\email{krain033@uottawa.ca} }
	
	\keywords{Uncertainty, Brain Lesions, MRI, Segmentation Sampling , Convolutional Neural Network, Discriminative}
	
	\abstract{Discriminative convolutional neural networks (CNNs), for which a voxel-wise conditional Multinoulli distribution is assumed, have performed well in many brain lesion segmentation tasks. For a trained discriminative CNN to be used in clinical practice, the patient's radiological features are inputted into the model, in which case a conditional distribution of segmentations is produced. Capturing the uncertainty of the predictions can be useful in deciding whether to abandon a model, or choose amongst competing models. In practice, however, we never know the ground truth segmentation, and therefore can never know the true model variance. In this work, segmentation sampling on discriminative CNNs is used to assess a trained model's robustness by analyzing the inter-sample Dice distribution on a new patient solely based on their magnetic resonance (MR) images. Furthermore, by demonstrating the inter-sample Dice observations are independent and identically distributed with a finite mean and variance under certain conditions, a rigorous confidence based decision rule is proposed to decide whether to reject or accept a CNN model for a particular patient. Applied to the ISLES 2015 (SISS) dataset, the model identified $7$ predictions as non-robust, and the average Dice coefficient calculated on the remaining brains improved by 12 percent. }
	
	\onecolumn \maketitle \normalsize \setcounter{footnote}{0} \vfill
	
	\section{\uppercase{Introduction}}
	\label{sec:introduction}
	
	\noindent Discriminative CNNs, such as those constructed by \cite{kamnitsas2017efficient,havaei2017brain,ronneberger2015u}, have consistently ranked on the top of the leaderboard in many brain lesion segmentation challenges \cite{maier2017isles,winzeck2018isles,bakas2018identifying}. Commonly, they are formulated by assuming a voxel-wise Multinoulli distribution, conditional on the MRI intensities of neighboring voxels where the parameters are obtained and estimated with a CNN. For the case of $2$ labels, if $x_j$ denotes the 3D MRI intensities used in the prediction of voxel $j$ over a discrete domain $\Omega \subset \mathbb{R}^3$, $Y_j$ represents the random label at voxel $j$, and $\Theta$ are the parameters of a CNN architecture, then the model is formulated as
	
	\begin{equation}
	Y_j|x_j \sim Bernoulli(\pi_j = CNN(\Theta,x_j)). \label{bern}  
	\end{equation}
	
	\noindent At prediction time, the translational invariance property of the CNN allows for a fast way to estimate all the voxel-wise conditional distributions, without having to separately feed their respective covariates. When a model is fully trained, the parameters in \eqref{bern} are replaced with an estimate: $\hat{\Theta}$. In this way, discriminative CNNs are capable of sampling segmentations by sampling the associated label at each voxel, conditionally independent of each other. Despite this source of variability, a decision rule of selecting the label with the highest probability is often applied rendering the segmentation deterministic.
	
	Measuring the performance variation of trained discriminative CNNs in brain lesion segmentation on a patient by patient basis can help clinical practitioners instill a degree of confidence in the use of automated segmentation methods. One metric of model performance is the Dice coefficient \cite{dice1945measures,sorensen1948method}, which measures the similarity between two 3D voxel-wise labeled images: $S_1$ and $S_2$. The Dice coefficient is a real number in $[0,1]$ given by
	
	\begin{equation}
	Dice(S_1,S_2) = \frac{2TP}{2TP + FN + FP},
	\label{dice}
	\end{equation}

	\noindent where a value of $0$ indicates no overlap and a value of $1$ indicates perfect similarity. TP, FP and FN refer to the number of true positive, false positive, and false negative voxels in the medical image respectively. Formulating this incentive mathematically, if $D$ denotes the Dice coefficient  random variable which incorporates randomness across patients, and across model segmentations, and $x$ collects all the covariates in \eqref{bern} across all voxels in the image for the patient to from a high dimensional vector, then the interest lies in measuring the model variance
	
	\begin{equation}
	Var(D|x,\hat{\Theta}). 
	\label{condvar}
	\end{equation}
	
	\noindent Here the Dice random variable is theoretically measured on the true segmentation, which is never known, and the random segmentations produced by a trained discriminative CNN model on the particular patient. Intuitively, the observations of $D$ can be generated by: observing a new random patient whose radiological features are drawn independently from other patients, randomly generating a segmentation from the trained discriminative CNN given their radiological features, and calculating the Dice coefficient against the ground truth. In \eqref{condvar}, the first source of variation is eliminated by conditioning on the patient. Estimating \eqref{condvar} can also help suggest the complete abandonment of an algorithm in favor of another, as some CNNs may be better tuned to deal with specific brain lesion characteristics like small lesion volume.

	Despite these advantages, some brain lesion segmentation challenges compute variability in the Dice metric across patient predictions \cite{maier2017isles}. Moreover, for a particular patient, only a single deterministic segmentation is produced by some decision rule. That is, $Var(D|\hat{\Theta})$ becomes the measure to be estimated, but this estimate does not solely account for model variability. It is important to note this measure is unconditional on the patient's radiological features, and thus inherently incorporates additional variability. The merit of this metric is in its applicability, as not all segmentation algorithm are capable of sampling segmentations. However in the case of discriminative CNNs, by its very nature, this metric can solely be used to quantify variation in the Dice coefficient across patients and could completely discard a model that may in fact perform well on certain kinds of patients. Another disadvantage is that this metric is obtained from the training set, and then must be generalized to arbitrary cases, which may present significant differences. Analysis of model performance variance on a single patient \eqref{condvar}, conditional on their radiological features, eliminates these extra sources of variation, and can be more useful in clinical practice by providing specialized patient care.    
	
	One step towards this direction was taken in the BraTS 2019 challenge, which is an extension of \cite{bakas2018identifying}. In particular, voxel-wise measures of segmentation uncertainty ranging from $0$ to $100$ were calculated from 3D MRIs for all patients individually. Then, at specified thresholds, uncertain voxels were filtered out in the calculation of the Dice coefficient. Depending on the structure of uncertainties, this method can reward or penalize the Dice score, but this can never be confirmed in practice since we never know the ground truth segmentation.
	
	Segmentation sampling and consequent analysis of the inter-sample Dice distribution has been undertaken for generative models, for instance, by L{\^e} et al. \cite{le2016sampling}. In their work, they use a Gaussian process to produce segmentation samples based on a single expert manual segmentation of grade $4$ gliomas. The mean of the inter-sample Dice distribution and variability of the segmentations can be controlled by a single model parameter. The samples can then be used in radiotherapy planning by delivering radiation to certain voxels and avoiding dose to uncertain voxels, where perhaps there are more sensitive tissues. 
	
	The contribution mentions the applicability of their method in evaluating uncertainty in the performance of segmentation algorithms, by repeatedly sampling segmentations off the ground truth and calculating the variability against a deterministic predicted segmentation. Though this method can be applied to arbitrary segmentation algorithms, the source of variation is not produced by the model predictions, as in the case of discriminative CNNs. As a consequence, the method is effective in assessing the effectiveness of a particular segmentation produced from a model, but not the model itself. Another related work by Roy et al. \cite{roy2018inherent}, shows that the inter-sample Dice coefficient correlates with Dice performance using a discriminative CNN for the segmentation of brain scans from children with psychiatric disorders (CANDI-13 dataset), and suggests it as a measure for quantifying uncertainty. 
	
	In this proposed work, the inter-sample Dice observations are shown to be independent and identically distributed samples from a distribution with finite variance and mean, under certain conditions. Segmentations are sampled directly from a discriminative CNN. The mean with a confidence interval of the inter-sample Dice distribution is then estimated by the central limit theorem and used in place of \eqref{condvar} to decide on whether to reject or accept the CNN model on patients with ischemic stroke. This chapter is organized as follows: Section 2 describes the methods, Section 3 presents the results, and a discussion follows in Section 4. 
	
	\section{\uppercase{Methods}}
	
\subsection{Architecture}

The CNN architecture considered is Wider2DSeg, which was originally constructed by Kamnitsas et al. \cite{kamnitsas2017efficient} and is a two dimensional variant of their 3D deepMedic architecture. Raina et al. \cite{bioimaging20} tuned this architecture in conjunction with additional symmetry covariates obtained from a reflective registration step to yield a $0.62$ Dice coefficient over a 7-fold cross validation on the ISLES 2015 SISS dataset. All the implementation details in this work are exactly as in the Wider2DSeg implementation of Raina et al. \cite{bioimaging20} with nonlinear symmetry covariates.

\subsection{Segmentation Sampling}

Referencing equation \eqref{bern}, suppose there are $V$ voxels in an image. The attention is restricted to the case of $2$ voxel labels, though this analysis can extend to an arbitrary finite number of labels. Let $S \in 
\{0,1\}^V$  represent the segmentation random vector of a trained discriminative CNN model, where each element $S_k = Y_j$, for unique $j \in \Omega$. The segmentation vector is unconditional and incorporates randomness across patients. Although in what follows it does not matter which permutation in the assignment of voxels to array elements is used, the assignment must be chosen and fixed.  Then, denote $x$ as the patient's radiological features as formulated in equation \eqref{condvar}, and $\hat{\Theta}$ as the estimated parameters of the CNN. Segmentations can be sampled conditional on a patient's radiological features if conditional independence is assumed across voxels. In this case, the distribution is completely known by

\begin{equation}
S|x \sim p(s;x,\hat{\Theta}) = \prod_{j=1}^{V}\pi_j(x_j,\hat{\Theta})^{s_j}(1-\pi_j(x_j,\hat{\Theta}))^{1-s_j}.
\label{conddist}
\end{equation}

\subsection{Inter-sample Dice Distribution}

Let $S^*_1$ and $S^*_2$ represent two iid model segmentation random vectors for a given patient, each distributed according to \eqref{conddist}. By independence, the joint distribution of these samples is given by

\begin{equation}
(S^*_1,S^*_2) \sim p(s^*_1;x,\hat{\Theta})p(s^*_2;x,\hat{\Theta}).
\label{jointdist}
\end{equation}

\noindent In this way, the inter-sample Dice random variable is just a function of the random vectors. In particular, if $S^*_2$ is seen as the ground truth, then the Dice coefficient of $S^*_1$ can be computed against it, and vice-versa. Define $\Gamma$ to be the inter-sample Dice random variable. Then

\begin{equation}
\Gamma = Dice(S^*_1,S^*_2).
\label{isd}
\end{equation}

\noindent In fact, the complete distribution of $\Gamma$ can be obtained, but can be quite computationally expensive as the number of voxels are often large. That is, if $\mathcal{D}_{(\cdot)}$ represents the support of a random variable, then $|\mathcal{D}_{S^*_1}| \leq 2^V$, where the upper bound is obtained by having $0 < \pi_j < 1$ for all voxels in the image. By independence of $S^*_1$ and $S^*_2$, this further implies that $|\mathcal{D}_{(S^*_1,S^*_2)}| \leq 2^{2V}$. Subsequently, after applying the Dice transformation \eqref{isd} which may at most associate a unique Dice value for each element in $\mathcal{D}_{(S^*_1,S^*_2)}$, the inequality $|\mathcal{D}_{\Gamma}| \leq 2^{2V}$ is obtained. Hence, computing the exact Dice distribution may require the calculation of at most $2^{2V}$ probabilities, where $V$ tends to be in the millions. Though, since $\mathcal{D}_{\Gamma}$ is bounded by $2^{2V}$, $\Gamma$ has a discrete distribution with a finite and countable support.

In order to conduct statistical inference on $E[\Gamma]$ it must also be finite, however, this is not the case. Notice that if $(S^*_1,S^*_2) = (0,0)$, $\Gamma$ is undefined, and this outcome always has a non-zero probability. In fact, this is the only way for $\Gamma$ to be undefined. To correct this, two necessary modifications are introduced: 1) $\exists$ $\pi_j \neq 0$, for some $j$ and 2) The inter-sample Dice random variable is redefined as

\begin{equation}
\Gamma^* = \Gamma|(S^*_1,S^*_2) \neq (0,0).
\end{equation}

\noindent The first condition ensures that $(0,0)$ can never be the only sample generated, thereby permitting its removal and still retaining a distribution through the redefinition in the second condition. The second condition redefines the inter-sample Dice random variable to be conditional on all samples drawn from \eqref{jointdist} except $(0,0)$. Now, $\forall \gamma \in \mathcal{D}_{\Gamma^*}, 0 \leq \gamma \leq 1$, where $\mathcal{D}_{\Gamma^*} \neq \emptyset$. Then

\begin{equation}
E[(\Gamma^*)^2] < \infty.
\end{equation}

As a consequence, sampling observations from this distribution, and estimating its mean from a sample can be used. Independent and identically distributed pairs of segmentations can be sampled from \eqref{jointdist}, and can generate an iid sample of inter-sample Dice observations. To adjust for the second modification, any $(0,0)$ sample, or equivalently undefined Dice score, is removed in the sampling phase. Moreover, since the inter-sample Dice distribution has a finite mean and variance, the central limit theorem can be applied, and associated confidence intervals can be constructed.

Computationally, this method can be undertaken by first producing the estimated probability tensors. At each voxel in the probability tensor are the estimated label probabilities obtained from the features and CNN parameter estimates. Once this tensor is calculated, segmentations are sampled by sampling from the associated Multinoulli distribution and iterating over all voxels. Naturally, these will produce segmentation samples that can be appropriately viewed as such since the format of the probability tensors align with that of the actual image. 

\subsection{Decision Rule}

Let $\gamma_1,..,\gamma_n$ be an iid sample of realized inter-sample Dice coefficients for a given patient, and construct a $(1-\alpha) \%$ approximate confidence interval, based on the central limit theorem. For a specified threshold, reject the use of a discriminative CNN model on a patient if the confidence interval is entirely below the threshold. In this manner, the clinician can justify with $(1-\alpha) \%$ confidence that the true mean of the inter-sample Dice distribution is below the given threshold, and appropriate actions to reject the model in favor of another can be undertaken. 
	
	\section{EXPERIMENTS AND RESULTS}
	
	\subsection{Dataset}
	
	The discriminative CNN model is trained and evaluated on the ISLES2015 (SISS) training data, which consists of $28$ patients with sub-acute ischemic stroke. The radiological features $x$ for a patient are 4 MR sequences: FLAIR, DWI, T1 and T1-contrast. Each image has a total of $230 \times 230 \times 154$ voxels. At each voxel, there are only two possible labels classes: lesion or non-lesion.

	\subsection{Efficacy of Decision Rule}
	
	The proposed segmentation sampling based decision rule is applied to the results of Raina et al. \cite{bioimaging20}, which yielded an average Dice coefficient (predicted against ground truth) of $0.62$ over a 7-fold cross-validation from single deterministic segmentations obtained by selecting the label with the highest probability. For each fold in the cross-validation and for each brain in the validation fold, $30$ inter-sample Dice observations are sampled, and

\begin{figure}[!h]
	\centering
	\includegraphics[scale=0.41]{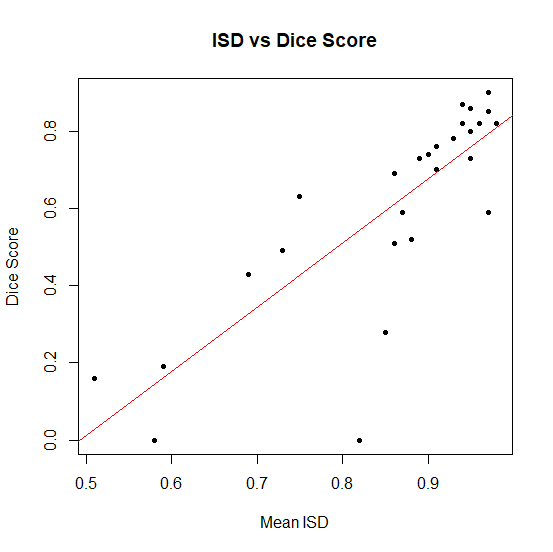}
	\caption{Plot of mean ISD against Dice score with regression line: $y = 1.66x - 0.82$, where $y$ is Dice score and $x$ is mean ISD. The F-test yielded a p-value of $1.4 \times 10^{-7}.$
	\label{fig:isdplot}} 
\end{figure}

	\begin{table}[!h]
		\centering
		\renewcommand{\arraystretch}{1.2}
		\vspace{0.3cm}
		\caption{Case by case hypothesis testing for Wider2dSeg on ISLES2015 SISS. The inter-sample Dice confidence interval is computed using the central limit theorem with $30$ samples. The bolded rows indicate rejecting the use of the CNN on the patient as per the decision rule.}\label{tab1}
		\begin{tabular}{|c|c|c|}
			\hline
			\bfseries Patient No. &  \bfseries ISD confidence & \bfseries Dice\\ 
			\hline
			1 &  0.940339 $\pm$ 0.000116 & 0.866798\\
			\hline
			2 & 0.955828 $\pm$ 0.000243 & 0.815299
			\\
			\hline
			3 & 0.897922 $\pm$ 0.001002 & 0.736231
			\\
			\hline
			4 & 0.946641 $\pm$ 0.000122 & 0.797607
			\\
			\hline
			5 & 0.944956 $\pm$ 0.000133 & 0.857821
			\\
			\hline
			6 & 0.965364 $\pm$ 0.000146 & 0.905630
			\\
			\hline
			7 &  0.943856 $\pm$ 0.000178 & 0.822416
			\\
			\hline
			8 & 0.909659 $\pm$ 0.000349 & 0.702697
			
			\\
			\hline
			9 & 0.969197 $\pm$ 0.000076 & 0.854602
			\\
			\hline
			10 &  0.875187 $\pm$ 0.000340 & 0.592438
			\\
			\hline
			11 & 0.933384 $\pm$ 0.000399 & 0.775896
			\\
			\hline
			12 &  0.888211 $\pm$ 0.000810 & 0.517170
			\\
			\hline
			13 &  0.850324 $\pm$  0.001449 & 0.277828
			\\
			\hline
			14 & 0.980536 $\pm$ 0.000095 & 0.815283
			\\
			\hline
			15 & 0.974488 $\pm$ 0.000179 & 0.887611
			\\
			\hline
			\bfseries 16 & \bfseries 0.580227 $\pm$ 0.004708 & \bfseries 0.009584
			
			\\
			\hline
			\bfseries 17 & \bfseries 0.508344 $\pm$ 0.005390 & \bfseries 0.164190
			\\
			\hline
			18 & 0.949571 $\pm$ 0.000435 & 0.730066
			\\
			\hline
			\bfseries 19 &  \bfseries 0.730039 $\pm$ 0.003967 & \bfseries 0.487031
			\\
			\hline
			20 & 0.950651 $\pm$ 0.000441 & 0.795393
			
			\\
			\hline
			
			\bfseries 21 & \bfseries 0.687765 $\pm$ 0.003033 & \bfseries 0.434674
			
			\\
			\hline
			22 &  0.863127 $\pm$ 0.000823 & 0.685439
			\\
			\hline
			\bfseries 23 &  \bfseries 0.754482 $\pm$ 0.002522 & \bfseries 0.634320
			\\
			\hline
			24 & 0.861986 $\pm$ 0.001278 & 0.507946
			\\
			\hline
			25 & 0.916227 $\pm$ 0.001130 & 0.761670
			\\
			\hline
			\bfseries 26 & \bfseries 0.598502 $\pm$ 0.006355 & \bfseries 0.191637
			
			\\
			\hline
			\bfseries 27 &  \bfseries 0.824157 $\pm$ 0.000633 & \bfseries 0.000156
			\\
			\hline
			28 & 0.887825 $\pm$ 0.001183 & 0.732581
			\\

			\hline
		\end{tabular}
	\end{table}

	\noindent the central limit theorem confidence interval is computed. Table \ref{tab1} displays the patient by patient results. The Pearson correlation between mean ISD and Dice over the $28$ patients was computed to be $r = 0.81$. Figure \ref{fig:isdplot} plots the mean ISD against Dice score for each patient, and depicts the fitted regression line. The threshold was set to $0.85$ with reference to the regression line in Figure \ref{fig:isdplot}, and corresponds to a Dice score of $0.60$. In addition, the confidence level is $95 \%$. By removing the rejected predictions, the average Dice coefficient increased from $0.62$ to $0.74$.

	\section{\uppercase{Discussion}}
	
\noindent The reason as to why ISD is highly correlated to Dice performance for discriminative CNNs, and can be used to detect weak segmentations is not entirely clear. As a counter-example, consider a model that always predicts $P(lesion) = 0$ for all but one voxel, which instead has $P(lesion) = 1$. Then, the ISD would always be $1$, since the model produces exactly the same segmentation with probability $1$, regardless of the input. However, it would be unexpected to see a high Dice score for this model. The coupling of CNN outputs and ISD for detecting uncertain segmentations requires further investigation for a deeper understanding of its performance.  

One important remark is that the Dice metric can be substituted by other metrics such as the sensitivity, specificity, mean squared error, or precision and the preceding analysis would also follow for these distributions, thereby permitting hypothesis testing. Moreover, the computations considered in this work were over the entire brain, but could also be calculated on specific regions of interest (ROIs). Deciding on which metrics to use, and applying them to more detailed brain sub-regions could improve the decision-making potential, and is a possible area of development.

Another point to remark is that the proposed method can be used to rigorously test competing discriminative models based on their respective inter-sample mean Dice confidence intervals, and select the most robust one on an individualized patient basis. Applying this unifying technique for all competing CNNs in a brain lesion challenge may exhibit the best possible performance, without any consideration to the ground truth. Segmentation challenges have recently begun to incorporate uncertainty analysis, but further work is required to apply these techniques on various types of brain lesion structures.

	\bibliographystyle{apalike}
	{\small
		\bibliography{biblio}}

\newcommand{\noopsort}[1]{} \newcommand{\printfirst}[2]{#1}
  \newcommand{\singleletter}[1]{#1} \newcommand{\switchargs}[2]{#2#1}
\begin{thebibliography}{}

\bibitem[Bakas et~al., 2018]{bakas2018identifying}
Bakas, S., Reyes, M., Jakab, A., Bauer, S., Rempfler, M., Crimi, A., Shinohara,
  R.~T., Berger, C., Ha, S.~M., Rozycki, M., et~al. (2018).
\newblock Identifying the best machine learning algorithms for brain tumor
  segmentation, progression assessment, and overall survival prediction in the
  brats challenge.
\newblock {\em arXiv preprint arXiv:1811.02629}.

\bibitem[Dice, 1945]{dice1945measures}
Dice, L.~R. (1945).
\newblock Measures of the amount of ecologic association between species.
\newblock {\em Ecology}, 26(3):297--302.

\bibitem[Havaei et~al., 2017]{havaei2017brain}
Havaei, M., Davy, A., Warde-Farley, D., Biard, A., Courville, A., Bengio, Y.,
  Pal, C., Jodoin, P.-M., and Larochelle, H. (2017).
\newblock Brain tumor segmentation with deep neural networks.
\newblock {\em Medical image analysis}, 35:18--31.

\bibitem[Kamnitsas et~al., 2017]{kamnitsas2017efficient}
Kamnitsas, K., Ledig, C., Newcombe, V.~F., Simpson, J.~P., Kane, A.~D., Menon,
  D.~K., Rueckert, D., and Glocker, B. (2017).
\newblock Efficient multi-scale {3D CNN} with fully connected {CRF} for
  accurate brain lesion segmentation.
\newblock {\em Medical image analysis}, 36:61--78.

\bibitem[L{\^e} et~al., 2016]{le2016sampling}
L{\^e}, M., Unkelbach, J., Ayache, N., and Delingette, H. (2016).
\newblock Sampling image segmentations for uncertainty quantification.
\newblock {\em Medical image analysis}, 34:42--51.

\bibitem[Maier et~al., 2017]{maier2017isles}
Maier, O., ~, B.~H., von~der Gablentz, J., H{\"a}ni, L., Heinrich, M.~P.,
  Liebrand, M., Winzeck, S., Basit, A., Bentley, P., Chen, L., et~al. (2017).
\newblock {ISLES 2015} - a public evaluation benchmark for ischemic stroke
  lesion segmentation from multispectral {MRI}.
\newblock {\em Medical image analysis}, 35:250--269.

\bibitem[Raina. et~al., 2020]{bioimaging20}
Raina., K., Yahorau., U., and Schmah., T. (2020).
\newblock Exploiting bilateral symmetry in brain lesion segmentation with
  reflective registration.
\newblock In {\em Proceedings of the 13th International Joint Conference on
  Biomedical Engineering Systems and Technologies - Volume 2: BIOIMAGING,},
  pages 116--122. INSTICC, SciTePress.

\bibitem[Ronneberger et~al., 2015]{ronneberger2015u}
Ronneberger, O., Fischer, P., and Brox, T. (2015).
\newblock U-net: Convolutional networks for biomedical image segmentation.
\newblock In {\em International Conference on Medical image computing and
  computer-assisted intervention}, pages 234--241. Springer.

\bibitem[Roy et~al., 2018]{roy2018inherent}
Roy, A.~G., Conjeti, S., Navab, N., and Wachinger, C. (2018).
\newblock Inherent brain segmentation quality control from fully convnet monte
  carlo sampling.
\newblock In {\em International Conference on Medical Image Computing and
  Computer-Assisted Intervention}, pages 664--672. Springer.

\bibitem[S{\o}rensen, 1948]{sorensen1948method}
S{\o}rensen, T.~J. (1948).
\newblock {\em A method of establishing groups of equal amplitude in plant
  sociology based on similarity of species content and its application to
  analyses of the vegetation on Danish commons}.
\newblock I kommission hos E. Munksgaard.

\bibitem[Winzeck et~al., 2018]{winzeck2018isles}
Winzeck, S., Hakim, A., McKinley, R., Pinto, J.~A., Alves, V., Silva, C.,
  Pisov, M., Krivov, E., Belyaev, M., Monteiro, M., et~al. (2018).
\newblock {ISLES} 2016 and 2017-benchmarking ischemic stroke lesion outcome
  prediction based on multispectral {MRI}.
\newblock {\em Frontiers in neurology}, 9.

\end{thebibliography}

\end{document}